\documentclass[aps,eqsecnum,preprint,floats,nofootinbib,letter]{revtex4}
\usepackage{epsfig}

\begin{document}
\def\bea{\begin{eqnarray}}
\def\eea{\end{eqnarray}}
\def\nc{N_c^{\rm eff}}
\def\vp{\varepsilon}
\def\drho{\bar\rho}
\def\deta{\bar\eta}
\def\a{{\cal A}}
\def\B{{\cal B}}
\def\c{{\cal C}}
\def\d{{\cal D}}
\def\e{{\cal E}}
\def\p{{\cal P}}
\def\t{{\cal T}}
\def\up{\uparrow}
\def\dw{\downarrow}
\def\vma{{_{V-A}}}
\def\vpa{{_{V+A}}}
\def\smp{{_{S-P}}}
\def\spp{{_{S+P}}}
\def\J{{J/\psi}}
\def\ov{\overline}
\def\Lqcd{{\Lambda_{\rm QCD}}}
\def\pr{{Phys. Rev.}~}
\def\prl{{Phys. Rev. Lett.}~}
\def\pl{{Phys. Lett.}~}
\def\np{{Nucl. Phys.}~}
\def\zp{{Z. Phys.}~}
\def\lsim{ {\ \lower-1.2pt\vbox{\hbox{\rlap{$<$}\lower5pt\vbox{\hbox{$\sim$}
}}}\ } }
\def\gsim{ {\ \lower-1.2pt\vbox{\hbox{\rlap{$>$}\lower5pt\vbox{\hbox{$\sim$}
}}}\ } }

\font\el=cmbx10 scaled \magstep2{\obeylines\hfill May, 2007}

\vskip 1.5 cm
\begin{center}{\large\bf Branching Ratios and $CP$ Asymmetries of
$B \to a_1(1260)~ \pi$ and $a_1(1260)~ K$ Decays}
\end{center}

\bigskip
\centerline{\bf Kwei-Chou Yang} \centerline{Department of Physics,
Chung Yuan Christian University} \centerline{Chung-Li, Taiwan 320,
Republic of China}
\bigskip
\bigskip
\bigskip
\bigskip
\centerline{\bf Abstract}
\bigskip
We present the studies of the decays $B\to a_1(1260)\, \pi$ and
$a_1(1260)\, K$ within the framework of QCD factorization. Due to
the G-parity, unlike the vector meson, the chiral-odd two-parton
light-cone distribution amplitudes of the $a_1$ are antisymmetric
under the exchange of quark and anti-quark momentum fractions in the
SU(2) limit. The branching ratios for $a_1 \, \pi$ modes are
sensitive to tree--penguin interference. The resultant ${\cal B}(B^0
\to a_1^\pm \pi^\mp)$ are in good agreement with the data. However,
using the current Cabibbo--Kobayashi--Maskawa angles,
$\beta=22.0^\circ$ and $\gamma=59.0^\circ$, our results for the
mixing-induced parameter $S$ and $\alpha_{\rm eff}$ differ from the
measurements of the time-dependent CP asymmetries in the decay
$B^0\to a_1^\pm \pi^\mp$ at about the $3.7\sigma$ level. This puzzle
may be resolved by using a larger $\gamma \gtrsim 80^\circ$. For
$a_1 K$ modes, the annihilation topologies give sizable
contributions and are sensitive to the first Gegenbauer moment of
the leading-twist tensor (chiral-odd) distribution amplitude of the
$a_1$ meson. The $B\to a_1 K$ amplitudes resemble the
corresponding $B\to \pi K$ ones very much. Taking the
ratios of corresponding CP-averaged $a_1 K$ and $\pi K$ branching
ratios, we can extract information relevant to the electroweak
penguins and annihilations. The existence of new-physics in the
electroweak penguin sector and final state interactions during
decays can thus be explored.


\pagebreak

\section{Introduction}

The first charmless hadronic $B$ decay involving a $1^3P_1$
axial-vector meson that has been observed is $B^0 \to a_1^\pm
(1260)\, \pi^\mp$
\cite{Aubert:2004xg,Aubert:2005xi,Abe:2005rf,Aubert:2006dd,Aubert:2006gb},
which goes through $b\to u \bar u d$. The measurements of
time-dependent CP asymmetries in hadronic $B$ decays originating
from $b\to u \bar u d$ can provide the information directly related
to the Cabibbo-Kobayashi-Maskawa (CKM) weak phase $\alpha\equiv\arg
(-V_{td} V_{tb}^*/V_{ud}V_{ub}^* )$ (or called $\phi_2$), for which
some results have been given from the data of $B\to \pi^+\pi^-,
\rho^\pm \pi^\mp$ and $\rho^\pm \rho^\mp$~\cite{hfag}. The BaBar
collaboration recently reported the observation of $B^0 \to a_1^\pm
(1260) \pi^\mp$, including $CP$ violating parameters, branching
fractions, and $\alpha_{\rm eff}$, where the bound on the difference
$\Delta \alpha=\alpha -\alpha_{\rm eff}$ can be constrained by using
the broken SU(3) flavor symmetry \cite{Gronau:2004tm,Gronau:2005kw}
or isospin analysis
\cite{Grossman:1997jr,Charles:1998qx,Lipkin:1991st}.

In this paper, we present the phenomenological studies of $B\to a_1
\pi$ and $a_1 K$ within the framework of QCD factorization, where
the former processes are tree-dominated, while the latter are
penguin-dominated. The $a_1(1260)$, which will be denoted by $a_1$
for simplicity, is the $1^3P_1$ state. Due to the G-parity, the
chiral-even two-parton light-cone distribution amplitudes (LCDAs) of
the $a_1$ are symmetric under the exchange of $quark$ and
$anti$-$quark$ momentum fractions in the SU(2) limit, whereas,
unlike the vector meson, the chiral-odd two-parton LCDAs are
antisymmetric. Ref.~\cite{Yang:2007zt} is the only literature so far
for the calculation of LCDAs of $1^3P_1$ axial-vector mesons. The
large first Gegenbauer moment of the leading-twist tensor
distribution amplitude of the $a_1$ meson \cite{Yang:2007zt} could
have a sizable impact on the annihilation amplitudes. On the other
hand, it is interesting to note that, for the axial-vector mesons
with quantum number $1^1P_1$, their chiral-even LCDAs are
anti-symmetric under the exchange of $quark$ and $anti$-$quark$
momentum fractions in the SU(3) limit, while the chiral-odd
two-parton LCDAs are symmetric \cite{Yang:2007zt,Yang:2005gk}. The
hadronic $B$ decays involving such a meson are sensitive to the
new-physics search \cite{Yang:2005tv,Das:2004hq}.

Because the axial-vector and pseudoscalar penguin contributions
interfere constructively in the dominant decay amplitudes of $\overline B\to
a_1 \overline K$, for which the $\overline K$ is emitted and $a_1$ shares the same
spectator quark within the $\overline B$ meson, the $\overline B\to a_1 \overline K$
amplitudes resemble very much the corresponding $\overline B\to \pi
\overline K$ ones. Moreover, larger CP asymmetries could be expected in the
$a_1^0 K^-$ and $a_1^0 \overline K^0$ modes due to the much lager
decay constant of the $a_1(1260)$, as compared with $\pi \overline K$
channels.

To resolve the puzzle about the observations of the decays $B\to \pi
K$ and $\pi \pi$ within the Standard Model (SM) \cite{hfag}, some
approaches were proposed, including considerations of final state
interactions (FSIs) \cite{Hou:1999st,Chua:2002wk,Cheng:2004ru}, and
use of SU(3) flavor symmetry to extract hadronic parameters from the
$\pi\pi$ data and then to predict $K\pi$ channels
\cite{Buras:2003dj,Buras:2004ub,Fleischer:2007mq}. On the other
hand, it was argued that new-physics with a large CP-violating phase
may exist in the electroweak penguin sector
\cite{Buras:2003dj,Buras:2004ub,Baek:2007yy}. The present studies
for $B\to a_1 \pi$ and $a_1 K$ modes can offer further tests for the
above theories.

The layout of the present paper is as follows. In Sec. II, we
discuss light-cone distribution amplitudes for an axial-vector
meson. A brief description for applying QCD factorization to the
decays $B\to a_1 \pi$ and $a_1 K$ is given in Sec. III, where some
relevant formulas are collected in Appendices~\ref{app:ai} and
\ref{app:A}. In terms of the notations $\alpha_i^p$ and $\beta_i^p$,
which were given in Ref.~\cite{BN}, one can find that the amplitudes for $AP$ modes have
the same expressions with those for $PP$ and $VP$ modes
(where $A\equiv$ the axial-vector meson, $P\equiv$ the pseudoscalar meson,
and $A\equiv$ the vector meson).
Sec. IV contains the numerical analysis for
the branching ratios and CP asymmetries. Our conclusions are
summarized in Sec. V.

\section{Two-parton LCDAs of the $a_1$ and Projection operators on the light-cone} \label{sec:DA}

For decays involving an axial-vector meson (denoted as $A$) in the
final state, the QCD corrections can turn the local quark-antiquark
operators into a series of nonlocal operators as
 \begin{eqnarray}\label{eq:nonlocal}
  &&  \langle A(P,\lambda)|\bar q_{1\,\alpha}(y) \, q_{2\,\delta}(x)|0\rangle
= -\frac{i}{4} \, \int_0^1 du \,  e^{i (u \, p y +
    \bar u p x)}
 \Bigg\{ f_A m_A \Bigg(
    \not\! p \gamma_5 \, \frac{\epsilon^{*(\lambda)}z}{pz} \,
    \Phi_\parallel(u)
  + \not\! \epsilon^{*(\lambda)}_{\perp} \gamma_5 \,
    g_\perp^{(a)}(u)
  \nonumber\\
  && ~~~  \, +  \epsilon_{\mu\nu\rho\sigma} \,
    \epsilon_{(\lambda)}^{*\nu}  p^{\rho} z^\sigma \, \gamma^\mu
    \, \frac{g_\perp^{(v)}(u)}{4}\Bigg)
  - \,f^{\perp}_A \Bigg(\not\!\epsilon_\perp^{*(\lambda)} \not\! p \gamma_5\,
  \Phi_\perp(u)
 - i \, \frac{m_A^2\,\epsilon^{*(\lambda)}z}{(p z)^2}
  \, \sigma_{\mu\nu}\gamma_5 \,p^{\mu} z^\nu
  \, h_\parallel^{(t)}(u)
  \nonumber \\
  && ~~~\, - i \, m_A^2 \, (\epsilon^{*(\lambda)} z) \,
  \frac{h_\parallel^{(p)}(u)}{2} \Bigg)
 \Bigg\}_{\delta\alpha}\!\!,
 \end{eqnarray}
where the chiral-even LCDAs are given by
\begin{eqnarray}
  &&\langle A(P,\lambda)|\bar q_1(y) \gamma_\mu \gamma_5 q_2(x)|0\rangle
  = if_{A} m_{A} \, \int_0^1
      du \,  e^{i (u \, p y + \bar u p x)}
   \left\{p_\mu \,
    \frac{\epsilon^{*(\lambda)} z}{p z} \, \Phi_\parallel(u)
         +\epsilon_{\perp\, \mu}^{*(\lambda)} \, g_\perp^{(a)}(u)
         \right\}, \nonumber\\ \label{eq:evendef1} \\
  &&\langle A(P,\lambda)|\bar q_1(y) \gamma_\mu
  q_2(x)|0\rangle
  = - i f_A m_A \,\epsilon_{\mu\nu\rho\sigma} \,
      \epsilon^{*\nu}_{(\lambda)} p^{\rho} z^\sigma \,
    \int_0^1 du \,  e^{i (u \, p y +
    \bar u p x)} \,
       \frac{g_\perp^{(v)}(u)}{4}, \label{eq:evendef2}
\end{eqnarray}
with $u \, (\bar u=1-u)$ being the momentum fraction carried by $q_1
(\bar q_2)$, and the chiral-odd LCDAs are given by
\begin{eqnarray}
  &&\langle A(P,\lambda)|\bar q_1(y) \sigma_{\mu\nu}\gamma_5 q_2(x)
            |0\rangle
  =  f_A^{\perp} \,\int_0^1 du \, e^{i (u \, p y +
    \bar u p x)} \,
\Bigg\{(\epsilon^{*(\lambda)}_{\perp\mu} p_{\nu} -
  \epsilon_{\perp\nu}^{*(\lambda)}  p_{\mu}) \,
  \Phi_\perp(u),\nonumber\\
&& \hspace*{+5cm}
  + \,\frac{m_A^2\,\epsilon^{*(\lambda)} z}{(p z)^2} \,
   (p_\mu z_\nu -
    p_\nu  z_\mu) \, h_\parallel^{(t)}(u)
\Bigg\},\label{eq:odddef1}\\
&&\langle A(P,\lambda)|\bar q_1(y) \gamma_5 q_2(x)
            |0\rangle
  =  f_A^{\perp} m_{A}^2 \epsilon^{*(\lambda)} z\,\int_0^1 du \, e^{i (u \, p y +
    \bar u p x)}  \, \frac{h_\parallel^{(p)}(u)}{2}\,.\label{eq:odddef2}
\end{eqnarray}
Here, throughout the present discussion, we define $z=y-x$ with
$z^2=0$, and introduce the light-like vector $p_\mu=P_\mu-m_A^2
z_\mu/(2 P z)$ with the meson's momentum ${P}^2=m_A^2$. Moreover,
the meson polarization vector $\epsilon^*_\mu$ has been decomposed
into longitudinal ($\epsilon^*_\parallel{}_\mu$) and transverse
($\epsilon^*_\perp{}_\mu$) {\it projections} defined as
\begin{eqnarray}\label{eq:polprojectiors}
 && \epsilon^*_\parallel{}_\mu \equiv
     \frac{\epsilon^* z}{P z} \left(
      P_\mu-\frac{m_A^2}{P z} \,z_\mu\right), \qquad
 \epsilon^*_\perp{}_\mu
        = \epsilon^*_\mu -\epsilon^*_\parallel{}_\mu\,,
\end{eqnarray}
respectively. The LCDAs $\Phi_\parallel, \Phi_\perp$ are of twist-2,
and $g_\perp^{(v)}, g_\perp^{(a)}, h_\perp^{(t)}, h_\parallel^{(p)}$
of twist-3. For the $a_1$ meson, due to G-parity, $\Phi_\parallel,
g_\perp^{(v)}$ and $g_\perp^{(a)}$ are symmetric with the
replacement of $u \leftrightarrow 1-u$, whereas $\Phi_\perp,
h_\parallel^{(t)}$ and $h_\parallel^{(p)}$ are antisymmetric in the
SU(2) limit \cite{Yang:2007zt}. Here, we restrict ourselves to
two-parton LCDAs with twist-3 accuracy.

Assuming that the axial-vector meson moves along the negative $z$-axis, the
derivation for the light-cone projection operator of an axial-vector
meson in the momentum space is in complete analogy to the case of
the vector meson. We separate the longitudinal and transverse parts
for the projection operator:
\begin{equation}
  M_{\delta\alpha}^A =  M_{\delta\alpha}^A{}_\parallel +
   M_{\delta\alpha}^A{}_\perp\,,
\label{rhomeson2}
\end{equation}
where only the longitudinal part is relevant in the present study
and given by
 \begin{eqnarray}
M^A_\parallel &=& -i\frac{f_A}{4} \, \frac{m_A(\epsilon^* n_+)}{2}
 \not\! n_- \gamma_5\,\Phi_\parallel(u)
-\frac{if_A^\perp m_A}{4}  \,\frac{m_A(\epsilon^* n_+)}{2E}
 \, \Bigg\{\frac{i}{2}\,\sigma_{\mu\nu}\gamma_5 \,  n_-^\mu  n_+^\nu \,
 h_\parallel^{(t)}(u)
\nonumber\\
&& \hspace*{-0.0cm} + \,i E\int_0^u dv \,(\Phi_\perp(v) -
h_\parallel^{(t)}(v)) \
     \sigma_{\mu\nu} \gamma_5 n_-^\mu
     \, \frac{\partial}{\partial k_\perp{}_\nu}
  - \gamma_5\frac{h_\parallel'{}^{(p)}(u)}{2}\Bigg\}\, \Bigg|_{k=u
  p}\,,
\end{eqnarray}
with the momentum of the quark $q_1$ in the $A$ meson being
\begin{eqnarray}
k_1^\mu = u E n_-^\mu +k_\perp^\mu + \frac{k_\perp^2}{4
uE}n_+^\mu\,,
\end{eqnarray}
for which $E$ is the energy of the axial-vector meson and the term proportional to $k_\perp^2$ is negligible. Here, for simplicity, we introduce two light-like vectors
$n_-^\mu\equiv (1,0,0,-1)$, and $n_+^\mu\equiv (1,0,0,1)$. In
general, the QCD factorization amplitudes can be written in terms of
the form $\int_0^1 du \, {\rm Tr} (M^A_\parallel \cdots)$.

In the following, we will give a brief summary for LCDAs of the
$a_1$ mesons, for which the detailed properties can be found in
Ref.~\cite{Yang:2007zt}. $\Phi_{\parallel,\perp}^{a_1}(u)$ can be
expanded in Gegenbauer polynomials:
\begin{eqnarray}\label{eq:t2-DA}
\Phi_{\parallel (\perp)}^{a_1}(u)= 6u\bar u \bigg[\sum_{i=0}^\infty
a_i^{\parallel\, (\perp),a_1} C_i^{3/2}(2u-1)\bigg]\,.
\end{eqnarray}
For the $\Phi_{\parallel (\perp)}^{a_1}(u)$, due to the G-parity,
only terms with even (odd) Gegenbauer moments survive in the SU(2)
limit. In the present work, we consider the approximations:
\begin{eqnarray}
\Phi_\parallel^{a_1} (u)& = & 6 u \bar u
 \left\{ 1 + a_2^{\parallel,a_1}\, \frac{3}{2} \bigg[ 5 (2u-1)^2  - 1 \bigg]
 \right\}, \label{eq:t2-1}\\
 \Phi_\perp^{a_1} (u) & = & 18 a_1^{\perp,a_1}\, u \bar u (2u-1).
\label{eq:t2-2}
\end{eqnarray}
Note that we have defined $f_{a_1}^\perp=f_{a_1}$ since the product
$f_{a_1}^\perp a_1^{\perp,a_1}$ always appears together. Neglecting
the three-parton distributions and terms proportional to the light
quark masses, we can relate the twist-3 distribution amplitudes to
the twist-2 ones by Wandzura-Wilczek relations
\cite{Wandzura:1977qf,Yang:2007zt} and then obtain:
\begin{eqnarray}\label{eq:ww}
 && h_\parallel^{(t)}(v)= (2v-1)\Bigg[ \int_0^v
\frac{\Phi_\perp(u)}{\bar u}du -\int_v^1
\frac{\Phi_\perp(u)}{u}du \Bigg] \equiv (2v-1) \Phi_a(v)\,, \nonumber\\
 && h_\parallel^{\prime(p)}(v)= -2\Bigg[ \int_0^v
\frac{\Phi_\perp(u)}{\bar u}du -\int_v^1
\frac{\Phi_\perp(u)}{u}du \Bigg] \equiv -2 \Phi_a(v)\,, \nonumber\\
 &&\int_0^v du \big( \Phi_\perp (u) -h^{(t)}_\parallel
(u))= v\bar v\Bigg[ \int_0^v \frac{\Phi_\perp(u)}{\bar u}du
-\int_v^1 \frac{\Phi_\perp(u)}{u}du\Bigg] \equiv v \bar v \Phi_a(v)
\,.
\end{eqnarray}
The normalization conditions for LCDAs are
\begin{eqnarray}
 \int_0^1 du \Phi_\parallel(u) = 1, \qquad
 \int_0^1 du \Phi_\perp (u) = 0 \,,
 \label{eq:nom-t2}
\end{eqnarray}
\begin{eqnarray}
 \int_0^1 du h_\parallel^{(t)}(u) = 0\,, \qquad
 \int_0^1 du h_\parallel^{(p)}(u) = 0\,.
 \label{eq:nom-t3}
\end{eqnarray}

For the pseudoscalar meson ($P$) with the four-momentum $P_\mu$, the
light-cone projection operator in the momentum space reads
 \begin{eqnarray}
M^P &=& i\frac{f_P}{4} \, E \not\! n_- \gamma_5\,\Phi_P(u) +
\frac{if_P \mu_P}{4}  \, \Bigg\{\frac{i}{2}\,\sigma_{\mu\nu}\gamma_5
\,  n_-^\mu  n_+^\nu \, \frac{\phi_\sigma'(u)}{6}
\nonumber\\
&& \hspace*{-0.0cm} - \,i E \frac{\phi_\sigma}{6}
     \sigma_{\mu\nu} \gamma_5 n_-^\mu
     \, \frac{\partial}{\partial k_\perp{}_\nu}
  - \gamma_5\frac{\phi_p(u)}{2}\Bigg\}\, \Bigg|_{k=u p}\,,
\end{eqnarray}
where $\mu_P=m_P^2/(m_1+m_2)$ is proportional to the chiral
condensate (with $m_{1,2}$ the masses of quarks) and the approximate forms of LCDAs that we use are
\begin{eqnarray}
\Phi_P (u)& = & 6 u \bar u \left\{ 1 + 3 a_1^P (2u-1)  + a_2^P\,
\frac{3}{2} \bigg[ 5 (2u-1)^2  - 1 \bigg]
 \right\}, \nonumber\\
 \Phi_p (u) & = & 1, \qquad \frac{\Phi_\sigma (u)}{6}=u(1-u).
\label{eq:pseudoscalar-DA}
\end{eqnarray}

\section{Decay amplitudes}\label{sec:amplitudes}
Within the framework of QCD factorization, in general the effective
weak Hamiltonian matrix elements for $\overline B \to M_1 M_2$
decays can be expressed in the form \cite{BN}
\begin{equation}\label{fac}
 \langle M_1 M_2 |{\cal H}_{\rm eff}|\overline B\rangle
  \! =\! \frac{G_F}{\sqrt2}\!\! \sum_{p=u,c} \! \lambda_p\,
 \!\langle M_1 M_2 |\!{\cal T_A}^{p}\!+\!{\cal T_B}^p
 |\overline B\rangle \,,
\end{equation}
where $\lambda_p\equiv V_{pb}V_{pq}^*$ with $q\equiv d$ or $s$,
$M_{2}$ is the emitted meson, and $M_{1} $ shares the same spectator
quark within the $\overline B$ meson. Considering a generic
$b$-quark decay, ${\cal T_A}^{p}$ describe contributions from naive
factorization, vertex corrections, penguin contractions and
spectator scattering, whereas ${\cal T_B}^{p}$ contain the weak
annihilation topologies.

For $\overline B$ decay processes, the QCD factorization approach advocated in
\cite{BBNS,BBNS1} allows us to compute the nonfactorizable
corrections in the heavy quark limit since only hard interactions
between the $(\overline B M_1)$ system and $M_2$ survive in the $m_b\to\infty$
limit. Naive factorization is recovered in the heavy quark limit and
to the zeroth order of QCD corrections. In this approach, the LCDAs
play an essential role.
 In the present study using the notations $\alpha_i^p$ and $\beta_i^p$
given in Ref.~\cite{BN}, the amplitudes for $AP$ modes have
the same expressions with those for $PP$ and $VP$ modes; $\overline B\to a_1
\pi, \ a_1 \overline K$ decay amplitudes in terms of
$\alpha_i^p$ and $\beta_i^p$ can be obtained from $\overline B\to \rho \pi, \ \rho \overline K$ \cite{BN} by setting $\rho \to a_1$.
However, one should note that the determination of the relative signs
of the detailed amplitudes behind the coefficients $\alpha_i^p$ and $\beta_i^p$ is non-trivial.

\subsection{Decay amplitudes due to ${\cal T_A}^{p}$}\label{subsec:Ta}

In general, ${\cal T_A}^{p}$ can be expressed in terms of
$c\,\alpha_i^p(M_1 M_2)\,
   X^{(\bar B M_1, M_2)}$,
where $c$ contains factors of $\pm 1$ and $\pm
1/\sqrt{2}$ arising from flavor structures of final-state mesons,
$\alpha_i$ are functions of the Wilson coefficients (see
Eq.~(\ref{ais})), and
 \begin{eqnarray}
  X^{(\bar B A, P)}
 &=& \langle P(q)| (V-A)_\mu|0\rangle \langle A(p)| (V-A)^\mu|\overline B(p_B)\rangle
 \nonumber\\
 &=& - 2 i f_P m_{A} V_0^{BA}(q^2) (\epsilon_{(\lambda)}^* p_B)\,,\\
  X^{(\bar B P, A)}
 &=& \langle A(q)| (V-A)_\mu|0\rangle \langle P(p)| (V-A)^\mu|\overline B(p_B)\rangle
 \nonumber\\
 &=& - 2 i f_A m_{A} F_1^{BP}(q^2) (\epsilon_{(\lambda)}^* p_B)\,.
 \end{eqnarray}
Here the decay constants of the pseudoscalar meson $P$ and the
axial-vector meson $A$ are defined by \cite{Bauer:1986bm}
 \begin{eqnarray} \label{eq:decay-const}
 && \langle P(p)|\bar q_2\gamma_\mu\gamma_5 q_1|0\rangle=-if_P p_\mu, \qquad
 \langle A(p,\lambda)|\bar q_2\gamma_\mu\gamma_5 q_1|0\rangle=if_A \epsilon^{(\lambda)*}_\mu.
 \end{eqnarray}
The form factors for the $B\to A$ and $P$ transitions are defined as
\cite{Bauer:1986bm}
\begin{eqnarray}
\langle{A}(p, \lambda)|A_\mu|{\overline B} (p_B)\rangle
 &=& i \frac{2}{m_B + m_{A}} \epsilon_{\mu\nu\alpha\beta} \epsilon_{(\lambda)}^{*\nu}
 p_B^\alpha p^{\beta} A^{BA}(q^2),
\nonumber \\
 \langle A (p,\lambda)|V_\mu|{\overline B}(p_B)\rangle
 &=& - \left[(m_B + m_{A}) \epsilon^{(\lambda)*}_{\mu} V_1^{BA}(q^2)
 - (\epsilon^{(\lambda)*} p_B)
(p_B + p)_\mu \frac{V_2^{BA}(q^2)}{m_B + m_{A}}\right]
\nonumber \\
&& + 2 m_{A} \frac{\epsilon^{(\lambda)*}\cdot p_B}{q^2} q^\mu
\left[V_3^{BA}(q^2) - V_0^{BA}(q^2)\right], \nonumber\\
 \langle P(p)|V_\mu|\overline B(p_B)\rangle &=&
 \left[ (p_B+p)_\mu-{m_B^2-m_P^2\over q^2}\,q_ \mu\right]
 F_1^{BP}(q^2)+{m_B^2-m_P^2\over q^2}q_\mu\,F_0^{BP}(q^2), \nonumber \\
 \end{eqnarray}
where $q = p_B - p$, $V_3^{BA}(0) = V_0^{BA}(0)$,
$F_1^{BP}(0)=F_0^{BP}(0)$ and
 \begin{eqnarray}
 V_3^{BA}(q^2)= \frac{m_B + m_{A}}{2 m_{A}} V_1^{BA}(q^2) - \frac{m_B -
m_{A}}{2 m_{A}} V_2^{BA}(q^2).
 \end{eqnarray}
The coefficients of the flavor operators $\alpha_i^{p}$ can be
expressed in terms of $a_i^{p}$ as follows:
\begin{eqnarray}\label{ais}
   \alpha_1(M_1 M_2) &=& a_1 (M_1 M_2) \,, \nonumber\\
   \alpha_2 (M_1 M_2) &=& a_2 (M_1 M_2) \,, \nonumber\\
   \alpha_3^{p}(M_1 M_2) &=&
     a_3^{p}(M_1 M_2) - a_5^{p}(M_1 M_2)\,,  \nonumber\\
   \alpha_4^{p}(M_1 M_2) &=& \left\{
    \begin{array}{cl}
     a_4^{p}(M_1 M_2) + r_{\chi}^{M_2}\,a_6^{p}(M_1 M_2)
      & \quad \mbox{for~} M_1 M_2=A\, P, \\
     a_4^{p}(M_1 M_2) - r_{\chi}^{M_2}\,a_6^{p}(M_1 M_2)
      & \quad \mbox{for~} M_1 M_2=P\, A,
    \end{array}\right.\\
   \alpha_{3,\rm EW}^{p}(M_1 M_2) &=&
     a_9^{p}(M_1 M_2) - a_7^{p}(M_1 M_2)\,, \nonumber\\
   \alpha_{4,\rm EW}^{p}(M_1 M_2) &=& \left\{
    \begin{array}{cl}
     a_{10}^{p}(M_1 M_2) + r_{\chi}^{M_2}\,a_8^{p}(M_1 M_2)
      & \quad \mbox{for~} M_1 M_2=A\, P\,, \\
     a_{10}^{p}(M_1 M_2) - r_{\chi}^{M_2}\,a_8^{p}(M_1 M_2)
      & \quad \mbox{for~} M_1 M_2=P\, A\,,
     \end{array}\right.\nonumber
\end{eqnarray}
where
 \begin{eqnarray} \label{eq:rchiP}
 r_\chi^P(\mu)&=&{2m_P^2\over m_b(\mu)(m_2+m_1)(\mu)}\,, \nonumber\\
 r_\chi^{A}(\mu) &=& \frac{2
 m_{A}}{m_b(\mu)}\,.
 \end{eqnarray}
The effective parameters $\alpha_i^{p}$ in Eq.~(\ref{ais})
to next-to-leading order in $\alpha_s$ can be expressed in forms
of~\cite{BN}.
 \begin{eqnarray}\label{eq:ai}
 a_i^{p}(M_{1} M_{2})\! &=& \! \bigg(c_i+{c_{i\pm1}\over
 N_c}\bigg)N_i(M_2) \nonumber\\
 && +{c_{i\pm1}\over N_c}\,{C_F\alpha_s\over
 4\pi}\Big[V_i(M_{2})+{4\pi^2\over
 N_c}H_i(M_{1} M_{2})\Big]+P_i^{p}(M_{2}),{\hspace{0.5cm}}
 \end{eqnarray}
where $c_i$ are the Wilson coefficients, $p=u,c$,
$C_F=(N_c^2-1)/(2N_c)$ with $N_c=3$, the upper (lower) signs refer
to odd (even) $i$, $M_{2}$ is the emitted meson, $M_{1} $ shares the
same spectator quark within the $B$ meson, and
\begin{equation}\label{eq:loterms}
   N_i = \Bigg\{
   \begin{array}{ll}
    ~0 \, & \quad \mbox{for $i=6,8,$ and $M_2=a_1$,} \\
    ~1 \, & \quad \mbox{for the rest.}
   \end{array}
\end{equation}
$V_i(M_2)$ account for vertex corrections, $H_i(M_1M_2)$ for hard
spectator interactions with a hard gluon exchange between the
emitted meson and the spectator quark of the $\overline B$ meson and
$P_i(M_2)$ for penguin contractions. The detailed results for the
above quantities are collected in Appendix~\ref{app:ai}. Note that
in the present case, some relative signs change in $H_i$ as compared
with the $PP$ and $VP$ modes.

\subsection{Decay amplitudes due to ${\cal T_B}^{p}$ --- annihilation topologies}\label{sebsec:Tb}

The $\overline B\to A P$ amplitudes governed by the annihilation
topologies read
\begin{eqnarray}\label{eq:ann}
\frac{G_F}{\sqrt2}\, \sum_{p=u,c} \! \lambda_p\, \!\langle A P
|\!{\cal T_B}^{p} |\overline B\rangle &=& -i \frac{G_F}{\sqrt{2}}f_B
f_{A} f_{P}\sum_{p=u,c} \lambda_p
 \bigg[\sum_{i=1}^{4} e_i b_i
+ e _5 b_{\rm 3,EW} + e _6 b_{\rm 4,EW} \bigg] , \ \ \ \
\end{eqnarray}
where the coefficients $e_i$ are process-dependent and weak
annihilation contributions are parameterized as
 \begin{eqnarray}\label{eq:bi}
 b_1 &=& {C_F\over N_c^2}c_1A_1^i, \qquad\quad b_3={C_F\over
 N_c^2}\left[c_3A_1^i+c_5(A_3^i+A_3^f)+N_cc_6A_3^f\right], \nonumber \\
 b_2 &=& {C_F\over N_c^2}c_2A_1^i, \qquad\quad b_4={C_F\over
 N_c^2}\left[c_4A_1^i+c_6A_2^f\right], \nonumber \\
 b_{\rm 3,EW} &=& {C_F\over
 N_c^2}\left[c_9A_1^{i}+c_7(A_3^{i}+A_3^{f})+N_cc_8A_3^{i}\right],
 \nonumber \\
 b_{\rm 4,EW} &=& {C_F\over
 N_c^2}\left[c_{10}A_1^{i}+c_8A_2^{i}\right].
 \end{eqnarray}
The subscripts 1,2 and 3 of $A_n^{i,f}$ denote the annihilation
amplitudes induced from $(V-A)(V-A)$, $(V-A)(V+A)$ and $(S-P)(S+P)$
operators, respectively, and the superscripts $i$ and $f$ refer to
gluon emission from the initial and final-state quarks,
respectively. For decays $B\to AP$, the detailed expressions for $A_n^{i,f}$ are given in
Appendix~\ref{app:A}.
$ \beta_i^p (M_1 M_2)$ are defined as
$$ \beta_i^p (M_1 M_2) =\frac{-i f_B f_{M_1} f_{M_2}}{X^{(\overline B M_1,M_2)}}b_i^p \,.$$

\section{Numerical results}

\subsection{Input parameters}

In the numerical analysis, we use the next-to-leading Wilson
coefficients in the naive dimensional regularization (NDR) scheme \cite{Buras96}. The relevant
parameters are summarized in Table~\ref{tab:inputs}
\cite{PDG,CKMfitter,Ball:2004ye,kcy:FF,Ball:2006wn}. The value of $f_B$ that we use is consistent with
the lattice average \cite{Bona:2006ah}. The current value of $F^{B\pi}(0)$ becomes a little smaller,
and is more suitable to explain the $\pi\pi$ data \cite{hfag}. We use the
light-cone sum rule results for the $B\to \pi, K$ \cite{Ball:2004ye}
and $B\to a_1$ \cite{kcy:FF} transition form factors, for which the
momentum dependence is parametrized as \cite{Ball:2006jz}
 \begin{equation}
 f(q^2)=f(0)\Bigg(\frac{1}{1-q^2/m_{B^*}^2}
 +\frac{r_{BZ(Y)} q^2/m_{B^*}^2}{1-\alpha_{BZ(Y)}q^2/m_{B}^2}\Bigg),
 \end{equation}
where $m_{B^*}$ is the lowest-resonance in the corresponding
channel. Note that since the mass of the $a_1$ meson is not small,
we have, for instance, $[ F_1^{B\pi} (m_{a_1}^2) /F_1^{B\pi} (0)]^2 \simeq 1.2$.
It means that the $q^2$ dependence of $B \to \pi, K$
form factors cannot be ignored in the prediction. As for the $B \to a_1$ form factor,
its $q^2$ dependence can be negligible due to the small mass of pseudoscalar mesons.
However, to be consistency, I also consider its $q^2$ dependence in the analysis.
Our light-cone sum rule result for $V_0^{Ba_1}(0)$ is a
little larger than the previous QCD sum rule calculation,
$0.23\pm0.05$ \cite{Aliev:1999mx}. It is interesting to compare with
other quark model calculations in the literature. The magnitude of
$V_0^{Ba_1}(0)$ is about 0.13 and $1.02\sim 1.22$ in the quark model
calculations in Ref.~\cite{Cheng:2003sm} and
Refs.~\cite{Scora:1995ty,Deandrea:1998ww}, respectively. The
magnitude of the former is too small and the latter is too large if
using them to compute the branching ratios of $\overline B^0\to
a_1^\pm \pi^\mp$ and then comparing with the data. The values of the
Gegenbauer moments of leading-twist LCDAs for the $a_1$ meson are
quoted from Ref.~\cite{Yang:2007zt}. The integral of the $B$ meson
wave function is parameterized as \cite{BBNS}
 \begin{eqnarray}
 \int_0^1 \frac{d\rho}{1-\rho}\Phi_1^B(\rho) \equiv
 \frac{m_B}{\lambda_B}\,,
 \end{eqnarray}
where $1-\rho$ is the momentum fraction carried by the light
spectator quark in the $B$ meson. Here we use $\lambda_B ({\rm
1~GeV}) =(350\pm 100)$~MeV.

There are three independent renormalization scales for describing
the decay amplitudes. The corresponding scale will be specified as
follows: (i) the scale $\mu_v=m_b/2$ for loop diagrams contributing
to the vertex and penguin contributions to the hard-scattering
kernels, (ii) $\mu_H=\sqrt{\mu_v\Lambda_h}$ for hard spectator
scattering, and (iii) $\mu_A=\sqrt{\mu_v\Lambda_h}$ for the
annihilation with the hadronic scale $\Lambda_h\approx 500$~MeV. We
follow \cite{BBNS} to parameterize the endpoint divergences
$X_A\equiv\int^1_0 dx/\bar x$ and $X_H\equiv\int^1_0 dx/\bar x$ in
the annihilation and hard-spectator diagrams, respectively, as
 \begin{eqnarray} \label{eq:XA}
 X_{A(H)}=\ln\left({m_B\over \Lambda_h}\right)(1+\rho_{A(H)} e^{i\phi_{A(H)}}),
 \end{eqnarray}
with the unknown real parameters $\rho_A, \rho_H$ and $\phi_A,
\phi_H$. We adopt the moderate value
$\rho_{A,H}\leq 0.5$ and arbitrary strong phases $\phi_{A,H}$ with
$\rho_{A,H}=0$ by default, i.e., we assign a 50\% uncertainty to the default value of $X_{A(H)}$
(with $\rho_{A,H}=0)$ \cite{Cheng:2005nb,Cheng:2007}; with the allowed ranges of $\rho_{A,H}$, the theoretical
predictions for $\pi K$ modes are consistent with the data.
Note that the $a_1 K$ rates could be sensitive to the magnitude of $\rho_A$.

\subsection{Results}

We follow the standard convention for the direct CP asymmetry
 \begin{equation}
 A_{CP}(\bar f)\equiv
 \frac{{\cal B}(\overline B^0 \to \bar f)-{\cal B}(B^0 \to f)}
      {{\cal B}(\overline B^0 \to \bar f)+{\cal B}(B^0 \to f)}\,.
 \end{equation}
The branching ratios given in the present paper are CP-averaged and
simply denoted by ${\cal B}(\overline B\to f)$. The numerical
results for CP-averaged branching ratios and direct CP asymmetries
are summarized in Tables~\ref{tab:br-1} and \ref{tab:acp-1},
respectively. The results for  time-dependent CP parameters of the
decay $B(t) \to a_1^\pm \pi^\mp$ are shown in
Table~\ref{tab:timeCP}.

\subsubsection{$\overline B\to a_1 \pi$}

The decay of the $B^0$ meson to $a_1^\pm \pi^\mp$ was recently
measured by the BaBar and Belle groups
\cite{Aubert:2004xg,Aubert:2005xi,Abe:2005rf,Aubert:2006dd,Aubert:2006gb}.
A recent updated result by BaBar yields~\cite{Aubert:2006dd}
 \begin{equation} \label{eq:babar-1}
 {\cal B}(B^0\to a_1^\pm \pi^\mp \to \pi^\mp \pi^\pm \pi^\pm \pi^\mp)=
 (16.6\pm 1.9 \pm 1.5)\times 10^{-6}.
 \end{equation}
Assuming that ${\cal B}(a_1^\pm \to \pi^\mp \pi^\pm \pi^\pm)$ equals
to ${\cal B}(a_1^\pm \to \pi^\mp \pi^0 \pi^0)$ and ${\cal B}(a_1^\pm
\to (3\pi)^\pm)$ equals to 100\%, they have obtained
 \begin{equation} \label{eq:babar-2}
 {\cal B}(B^0\to a_1^\pm \pi^\mp)=
 (33.2\pm 3.8 \pm 3.0)\times 10^{-6}.
 \end{equation}
Very recently, the measurements of time-dependent CP asymmetries in
the decay $B^0 \to a_1^\pm \pi^\mp$ have been reported by the BaBar
collaboration~\cite{Aubert:2006gb}. From the measurements, the
individual branching ratios of $\overline B^0\to a_1^+ \pi^-$ and
$a_1^- \pi^+$ can be obtained. As given in Table~\ref{tab:br-1}, our
theoretical results are in good agreement with experiment. It was
shown in Ref.~\cite{Yang:2003sg} that three-parton Fock states of
$M_2$ can give non-small corrections to $\alpha_2^p$, so that
$|\alpha_2^p|\simeq 0.30$, If so, we can expect ${\cal B}(\overline
B^0 \to a_1^0 \pi^0)\gtrsim 1.6\times 10^{-6}$, which can be tested
in the future measurement.

The $\overline B\to a_1 \pi$ amplitudes are analogous to the
corresponding $\overline B\to \rho \pi$ ones \cite{Hou:1999tf}. The
tree(T)-penguin(P) interference depends on the sign of $\sin\gamma$
(where $V_{ub}=|V_{ub}|e^{-i\gamma}$) and the relative sign between
Re($\alpha_1^p$) and Re($\alpha_4^p$); for $\sin\gamma>0$, the T-P
interference is destructive in $\overline B^0\to a_1^\mp\pi^\pm, B^-
\to a_1^0\pi^-$, while it is constructive in $B^- \to a_1^-\pi^0$.
Because the amplitudes of $a_1 \pi$ and $\rho \pi$ modes are dominated by the terms
with $\alpha_{1}$ and $\alpha^p_{4}$, and ${\rm Re}[\alpha_4^p(\pi
a_1)] \approx {\rm Re}[\alpha_4^p(a_1 \pi)/3] \approx {\rm
Re}[\alpha_4^p(\pi \rho)] \approx - {\rm Re}[\alpha_4^p(\rho \pi)]
\approx -0.034$, one can easily obtain the following relations,
 \begin{eqnarray}\label{eq:ratio-1}
 \frac{{\cal B}(\overline B^0 \to a_1^- \pi^+)}
  {{\cal B}(\overline B^0 \to \rho^- \pi^+)} &\approx&
 \Bigg(
 \frac{F_1^{B\pi}(m_{a_1}^2) f_{a_1}}{F_1^{B\pi}(m_{\rho}^2) f_{\rho}}
 \Bigg)^2\,,\nonumber\\
 \frac{{\cal B}(\overline B^0 \to a_1^+ \pi^-)}
  {{\cal B}(\overline B^0 \to \rho^+ \pi^-)} & < &
 \Bigg(
 \frac{V_0^{Ba_1}(m_{\pi}^2)}{A_0^{B\rho}(m_{\pi}^2)}
 \Bigg)^2\,,\nonumber \\
 \frac{{\cal B}(B^- \to a_1^0 \pi^-)}
  {{\cal B}(B^- \to \rho^0 \pi^-)} & < &
 \Bigg(
 \frac{V_0^{Ba_1}(m_{\pi}^2)}{A_0^{B\rho}(m_{\pi}^2)}
 \Bigg)^2\,,\nonumber \\
 \frac{{\cal B}(B^- \to a_1^- \pi^0)}
  {{\cal B}(B^- \to \rho^- \pi^0)} & > &
 \Bigg(
 \frac{F_1^{B\pi}(m_{a_1}^2) f_{a_1}}{F_1^{B\pi}(m_{\rho}^2) f_{\rho}}
 \Bigg)^2 \approx  \frac{{\cal B}(\overline B^0 \to a_1^- \pi^+)}
  {{\cal B}(\overline B^0 \to \rho^- \pi^+)}\,,
 \end{eqnarray}
which can offer constraints on the magnitudes of
$f_{a_1}$ and $V_0^{Ba_1}(m_{\pi}^2)$. Moreover, the ratio
${\cal B}(\overline B^0 \to a_1^- \pi^+)/{\cal B}(\overline B^0 \to
a_1^+ \pi^-)$ is
\begin{eqnarray}\label{eq:ratio-2}
 \frac{{\cal B}(\overline B^0 \to a_1^- \pi^+)}
  {{\cal B}(\overline B^0 \to a_1^+ \pi^-)} &=&
 \Bigg(
 \frac{F_1^{B\pi}(m_{a_1}^2) f_{a_1}}{V_0^{B a_1}(m_{\pi}^2) f_{\pi}}
 \Bigg)^2 \Bigg\{ 1+ {\rm Re}\Bigg[\frac{\lambda_t}{\lambda_u}
  \Bigg(\frac{\alpha_4(\pi a_1) -\alpha_4(a_1 \pi)+\beta_3 (\pi a_1) -\beta_3(a_1\pi)}
  {\alpha_1(\pi a_1)}\Bigg)
  \nonumber\\
  & + & 2 \Bigg(\frac{V_0^{Ba_1}(m_{\pi}^2) f_\pi}{F_1^{B\pi}(m_{a_1}^2) f_{a_1}}-1\Bigg)
  {\rm Re}\Bigg[\frac{\beta_1(\pi a_1)}{\alpha_1(\pi
  a_1)}\Bigg]\Bigg\} + {\cal{O}}(\alpha^p_{4,{\rm EW}},\beta^p_{4},\beta^p_{3,{\rm EW}},\beta^p_{4,{\rm EW}})\,,
 \end{eqnarray}
which is not only sensitive to the form factor and decay constant of the $a_1$ meson
but also to the weak phase $\gamma$. The measurement of the above ratio allow us to obtain
the further constraint on the value of $\gamma$.

The large direct
CP asymmetries may result from the non-zero value of the weak
annihilation parameter ($\rho_A$) and its corresponding phase. See
Table~\ref{tab:acp-1}. With default parameters, the direct CP
asymmetries for $a_1^+\pi^-, a_1^-\pi^+, a_1^-\pi^0, a_1^0 \pi^-$
are only at a few percent level, whereas it can be very remarkable
for the $a_1^0 \pi^0$ mode. At the present time, the large errors
in the measurements for $CP$ asymmetries do not allow us
to draw any particular conclusion in comparison with theoretical predictions.
(See Tables~\ref{tab:acp-1} and \ref{tab:timeCP}.)
\begin{table}[t]
\centerline{\parbox{14cm}{\caption{\label{tab:inputs} Summary of
input parameters.}}} \vspace{0.1cm}
\begin{center}
{\tabcolsep=0.96cm\begin{tabular}{|c|c|c|c|c|} \hline\hline
\multicolumn{4}{|c|}{Running quark masses [GeV] and the strong coupling constant
\cite{BN,PDG}} \\
\hline   $m_c(m_c)$ & $m_s(2\,\mbox{GeV})$
       & $(m_u+m_d)/(2 m_s)$ & $\alpha_s(1~{\rm GeV})$ \\
\hline
 $1.3$ & $0.09\pm 0.01$ & $0.0413$ &  $0.497$\\
\hline
\end{tabular}}
{\tabcolsep=1.485cm\begin{tabular}{|c|c|c|c|} \hline
\multicolumn{4}{|c|}{Wolfenstein parameters for the CKM matrix elements \cite{CKMfitter}} \\
\hline  $A$   &  $\lambda$ &  $\bar\rho$ & $\bar\eta$ \\
\hline 0.806& $0.2272$ & $0.195$ & $0.326$  \\
\hline
\end{tabular}}
{\tabcolsep=1.445cm\begin{tabular}{|c|c|c|c|} \hline
\multicolumn{4}{|c|}{Decay constants for mesons [MeV] \cite{PDG,Cheng:2005nb,Yang:2007zt}} \\
\hline
$f_\pi$ & $f_K$ & $f_{B}$ & $f_{a_1}$   \\
\hline
131 & 160 & $195\pm 10$ & $238\pm 10$ \\
\hline
\end{tabular}}
{\tabcolsep=0.53cm\begin{tabular}{|cccc|} \hline
\multicolumn{4}{|c|}{Form factors and parameters for their $q^2$
dependence \cite{Ball:2004ye,kcy:FF}} \\
\hline $F_1^{B\pi}(0)=0.26 \pm 0.03$ & $\alpha_{BZ}=0.40$ & $r_{BZ}=0.64$ & $m_1=m_{B^*}=5.32$~GeV\\
\hline
$F_1^{BK}(0)=0.33\pm 0.04$ & $\alpha_{BZ}=0.95$ & $r_{BZ}=0.52$ & $m_1=m_{B^*_s}=5.41$~GeV\\
\hline \hline
$V_0^{Ba_1}(0)=0.28 \pm 0.03$ & $\alpha_{Y}=0.90$ & $r_{Y}=0.65$ & $m_1=m_{B^*}=5.32$~GeV\\
\hline
\end{tabular}}
{\tabcolsep=0.58cm\begin{tabular}{|c|c|c|c|c|} \hline
\multicolumn{5}{|c|}{Gegenbauer moments for leading-twist LCDAs of
mesons at scale 1~GeV
\cite{Ball:2006wn,Yang:2007zt}} \\
\hline
$a_2^\pi$ & $a_1^K$ & $a_2^K/a_2^\pi$ & $a_2^{\parallel,a_1}$ & $a_1^{\perp,a_1}$\\
\hline
$0.25\pm0.15$ & $0.06\pm0.03$ & $1.05\pm0.15$ & $-0.03\pm0.02$ & $-1.04\pm0.34$\\
\hline
\end{tabular}}
\end{center}
\end{table}

\begin{table}[t]
\caption{CP-averaged branching fractions for the decays $B\to
a_1(1260) \pi$ and $a_1(1260) K$ (in units of $10^{-6}$). The
theoretical errors correspond to the uncertainties due to variation
of (i) Gegenbauer moments, decay constants, (ii) quark masses, form
factors, and (iii) $\lambda_B, \rho_{A,H}$, $\phi_{A,H}$,
respectively, added in quadrature.}
 \label{tab:br-1}
\begin{ruledtabular}
\begin{tabular}{l r c c }
Mode & Theory & Expt. (BaBar) \cite{Aubert:2006dd,Aubert:2006gb} & Expt. (Belle) \cite{Abe:2005rf}\\
\hline
 $\overline B^0\to a_1^+ \pi^-$
 & $8.7^{+0.2+2.4+2.1}_{-0.2-2.0-1.3}$ & $12.2\pm 4.5$ \nonumber \\
 $\overline B^0\to a_1^- \pi^+$
 & $25.1^{+2.5+6.5+2.6}_{-2.4-5.8-1.6}$ & $21.0\pm5.4$\nonumber \\
 $\overline B^0\to a_1^\pm \pi^\mp$
 & $33.8^{+2.6+8.9+4.7}_{-2.6-7.8-2.9}$ & $33.2\pm 5.0$ & $48.6\pm 5.6$\nonumber \\
 $\overline B^0\to a_1^0 \pi^0$ & $0.7^{+0.1+0.2+0.7}_{-0.1-0.1-0.3}$ &
 \nonumber\\
 $B^-\to a_1^- \pi^0$ & $14.9^{+1.9+3.7+2.4}_{-1.7-3.3-2.1}$ & \nonumber \\
 $B^-\to a_1^0 \pi^-$ & $7.3^{+0.3+1.7+1.3}_{-0.3-1.5-0.9}$ & \nonumber \\
 \\
 $\overline B^0\to a_1^+ K^-$
 & $15.1^{+1.2+12.7+21.2}_{-1.2- ~6.3- ~\, 7.2}$ & \nonumber \\
 $\overline B^0\to a_1^0 \overline K^0$ & $6.0^{+0.4+5.6+9.7}_{-0.4-2.6-3.1}$ & \nonumber \\
 $B^-\to a_1^-  \overline K^0$ & $19.1^{+1.3+15.5+24.5}_{-1.3- ~ 7.8-11.0}$ & \nonumber \\
 $B^-\to a_1^0 K^-$ & $11.8^{+1.0+8.7+13.1}_{-1.0-4.6- ~ 4.8}$ & \nonumber
 \\
\end{tabular}
\end{ruledtabular}
\end{table}

\begin{table}[th]
\caption{Direct CP asymmetries for the decays $B\to a_1(1260)\, \pi$
and $a_1(1260)\, K$ (in \%). See Table~\ref{tab:br-1} for errors.}
 \label{tab:acp-1}
\begin{ruledtabular}
\begin{tabular}{l r r| l r}
Mode & Theory & BaBar \cite{Aubert:2006gb,hfag} & Mode & Theory \\
\hline
 $\overline B^0\to a_1^+ \pi^-$   &
 $-3.2^{+0.1+0.3+20.1}_{-0.0-0.5-19.5}$& $7 \pm 21\pm 15$
 & $\overline B^0\to a_1^+ K^-$   & $2.7^{+0.2+0.9+11.8}_{-0.2-0.8-11.9}$
 \nonumber \\
 $\overline B^0\to a_1^-\pi^+$    &
 $-1.7^{+0.1+0.1+13.6}_{-0.1-0.0-13.4}$  & $15\pm15\pm \, 7$
 &$\overline B^0\to a_1^0 \overline K^0$ & $-7.9^{+0.7+2.1+7.6}_{-0.7-2.2-8.3}$
 \nonumber \\
 $  \overline B^0\to a_1^0 \pi^0$  &
 $69.3^{+5.4+ 6.9+25.0}_{-6.1-8.9-74.7}$&
 &$B^-\to a_1^- \overline K^0$    & $0.7^{+0.0+0.1+0.6}_{-0.0-0.1-0.1}$
 \nonumber \\
 $B^-\to a_1^- \pi^0$  & $-0.4^{+0.4+0.2+11.1}_{-0.4-0.1-11.1}$
                                       &
 & $B^-\to a_1^0  K^-$             & $8.8^{+0.5+1.5+12.1}_{-0.5-1.7-13.4}$
 \nonumber\\
 $B^-\to a_1^0 \pi^-$ & $-0.5^{+0.5+1.5+13.0}_{-0.3-2.4-14.6}$&  & \nonumber \\
\end{tabular}
\end{ruledtabular}
\end{table}

\begin{table}[t]
\centerline{\parbox{14cm}{\caption{\label{tab:timeCP} Parameters of
the time-dependent $B\to a_1^\pm \pi^\mp$ decay rate asymmetries.
$S$ and $\Delta S$ are computed for $\beta=22.0^\circ$,
corresponding to $\sin(2\beta)=0.695$, and $\gamma =59.0^\circ$. See
Table~\ref{tab:br-1} for errors.}}} \vspace{0.0cm}
\begin{center}
\begin{ruledtabular}
\begin{tabular}{|c|c r|}
\multicolumn{1}{|c|}{} & Theory  & Experiment (BaBar) \cite{Aubert:2006gb}\\
\hline\hline
 $A_{\rm CP}^{a_1\pi}$ &
 $\phantom{-}0.01^{\,+0.00\,+0.00\,+0.05}_{\,-0.00\,-0.00\,-0.05}$
 & $-0.07\pm0.07\pm0.02$\\
 $C$ & $\phantom{-}0.02^{\,+0.00\,+0.00\,+0.13}_{\,-0.00\,-0.00\,-0.13}$ &
   $-0.10\pm0.15\pm0.09$ \\
$S$ &
 $ -0.55^{\,+0.02\,+0.04\,+0.08}_{\,-0.02\,-0.06\,-0.13}$ &
   $0.37\pm0.21\pm0.07$ \\
$\Delta C$ &
$\phantom{-}0.48^{\,+0.04\,+0.02\,+0.03}_{\,-0.04\,-0.04\,-0.05}$ &
   $0.26\pm0.15\pm0.07$ \\
 $\Delta S$ & $-0.01^{\,+0.00\,+0.00\,+0.03}_{\,-0.00\, -0.00\,-0.03}$ &
   $-0.14\pm0.21\pm0.06$ \\
 & & \\
 $\alpha_{\rm eff}^+$  & $(105.1^{\, +0.3\, +0.9\, +4.4}_{\,-0.3\, -0.5\, -2.4})^\circ$
 &  \\
 $\alpha_{\rm eff}^-$  & $(113.9^{\, +0.6\, +3.2\, +6.4}_{\,-0.6\, -2.1\, -3.6})^\circ$
 &  \\
 $\alpha_{\rm eff}$  & $(109.5^{\, +0.5\, +2.1\, +5.4}_{\,-0.5\, -1.3\, -3.0})^\circ$
 &  $(78.6\pm 7.3)^\circ$ \\
\end{tabular}
\end{ruledtabular}
\end{center}
\end{table}

\subsubsection{Time-dependent CP for $B(t)\to a_1^\pm \pi^\mp$}

\begin{figure}[t]
\centering \mbox{{\epsfig{figure=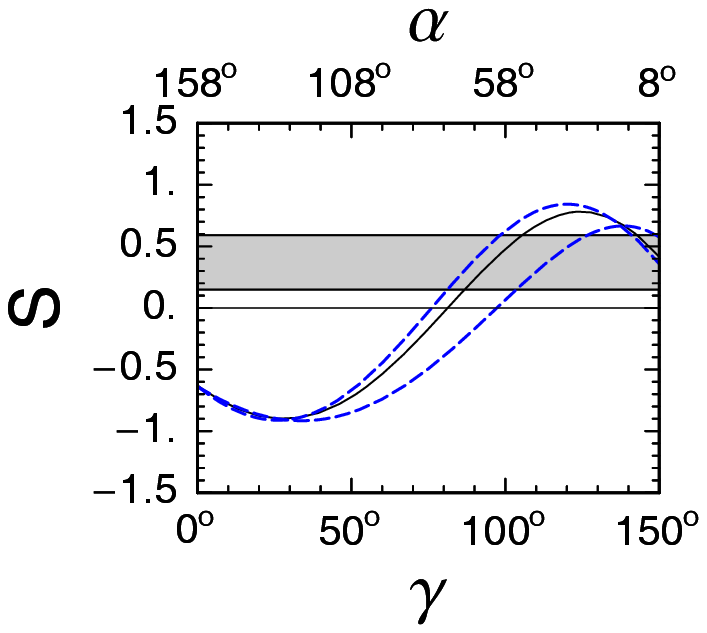,width=6.5cm} \hskip1cm
\epsfig{figure=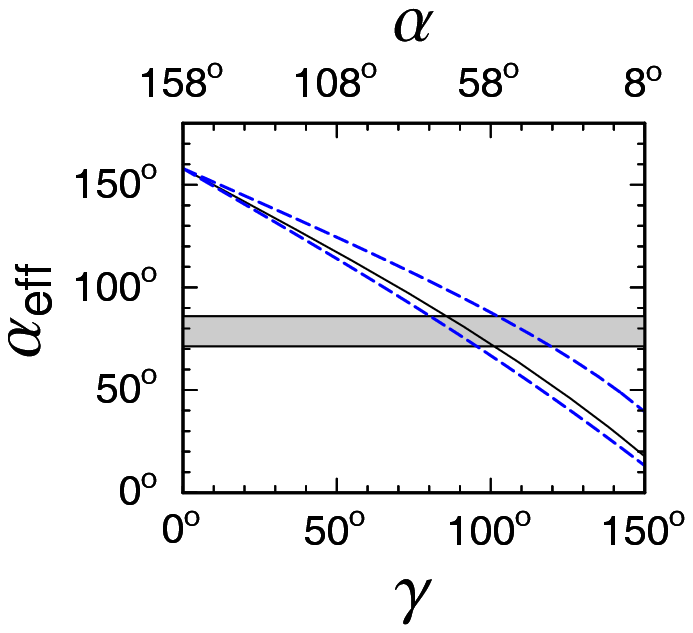,width=6.5cm}}} \caption{\label{fig:s}
$S$ and $\alpha_{\rm eff}$ versus $\gamma$ (and $\alpha$) for
adopting $\beta=22^\circ$. The solid curves are obtained by using
the central values (default values) of input parameters. The region
between two dashed lines is the theoretical variation within the
allowed range of input parameters.}
\end{figure}

Following Ref.~\cite{Gronau:2005kw}, we define
\begin{eqnarray}
 A_+ & \equiv & A(B^0\to a_1^+\pi^-)~,~~~
A_-  \equiv A(B^0\to a_1^-\pi^+)~,\nonumber\\
\overline{A}_+ & \equiv & A(\overline B^0\to a_1^-\pi^+)~,~~~
\overline{A}_- \equiv A(\overline B^0 \to a_1^+\pi^-)~.
\end{eqnarray}
Neglecting CP violation in the $B^0-\overline B^0$ mixing and the
width difference in the two $B^0$ mass eigenstates, time-dependent
decay rates for initially $B^0$ decaying into $a_1^\pm\pi^\mp$ can
be parameterized by
\begin{eqnarray}\label{eq:Gammat}
\Gamma(B^0(t) \to a_1^\pm\pi^\mp) &=& e^{-\Gamma t}
\frac{1}{2}\left (|A_{\pm}|^2 + |\overline{A}_{\mp}|^2\right)\nonumber\\
 && \times \Big[1 +(C \pm \Delta C)\cos\Delta mt - (S \pm \Delta S)\sin\Delta mt\Big]~,
\end{eqnarray} where
\begin{equation}\label{CSdef} C \pm \Delta C \equiv \frac{|A_{\pm}|^2
- |\overline{A}_{\mp}|^2}{|A_{\pm}|^2 + |\overline{A}_{\mp}|^2}~,
\end{equation}
 and
\begin{equation} S \pm \Delta S \equiv \frac{2{\rm
Im}(e^{-2i\beta}\overline{A}_{\mp} A^*_{\pm})} {|A_{\pm}|^2 +
|\overline{A}_{\mp}|^2}~.
\end{equation}
Here $\Delta m$ denotes the neutral $B$ mass difference and $\Gamma$
is the average $B^0$ width. For an initial $\bar B^0$ the signs of
the $\cos \Delta mt$ and $\sin \Delta mt$ terms are reversed. The
four decay modes define five asymmetries: $C, S, \Delta C, \Delta
S$, and the overall CP violating $A_{CP}^{a_1\pi}$,
 \begin{equation}
 A_{CP}^{a_1\pi} \equiv
 \frac{|A_+|^2 + |\overline A_-|^2 - |A_-|^2 - |\overline A_+|^2}
 {|A_+|^2 + |\overline A_-|^2 + |A_-|^2 + |\overline A_+|^2}\,.
 \end{equation}
Two $\alpha$-related phases can be defined by
 \begin{eqnarray}
 \alpha_{\rm eff}^\pm \equiv \frac{1}{2}\arg (e^{-2i\beta} \overline
 A_\pm A_\pm^*) \,,
 \end{eqnarray}
which coincide with $\alpha$ in the limit of vanishing penguin
amplitudes. The average of $\alpha_{\rm eff}^+$ and $\alpha_{\rm
eff}^-$ is called $\alpha_{\rm eff}$:
 \begin{eqnarray}\label{eq:alpha-eff}
 \alpha_{\rm eff}
 &\equiv& \frac{\alpha^+_{\rm eff} + \alpha^-_{\rm eff}}{2}
 = \frac{1}{4} \Bigg[ \arcsin \Bigg(\frac{S + \Delta S}
    {\sqrt{1- (C + \Delta C)^2}}\Bigg)
 + \arcsin\Bigg(\frac{S - \Delta S}{\sqrt{1- (C - \Delta C)^2}}\Bigg) \Bigg].
 \ \ \ \ \ \ \
\end{eqnarray}
The numerical results for the time-dependent CP parameters are
collected in Table~\ref{tab:timeCP}. The magnitudes of
$A_{CP}^{a_1\pi}, C$ and $\Delta S$ are small in the QCD
factorization calculation, where $C$ is sensitive to the
annihilations and can be $\sim 10\%$ in magnitude. $\Delta C$
describes the asymmetry between ${\cal B}(B^0 \to a_1^+ \pi^-) +
{\cal B}(\overline B^0 \to a_1^- \pi^+)$ and ${\cal B}(B^0 \to a_1^-
\pi^+) + {\cal B}(\overline B^0 \to a_1^+ \pi^-)$, and thus can be
read directly from Tables~\ref{tab:br-1} and \ref{tab:acp-1}.
Neglecting penguin contributions, $S$ and $\alpha_{\rm eff}$, which
depend on $\alpha (=\pi-\beta-\gamma)$, coincides with $\sin
2\alpha$ and $\alpha$, respectively, in the SM. Using
$\alpha=99.0^\circ$, i.e., $\gamma=59.0^\circ$, the numerical
results for $S$ and $\alpha_{\rm eff}$ differ from the experimental
values at about the $3.7\sigma$ level. This puzzle may be resolved
by using a smaller $\alpha=\pi -\beta - \gamma\lesssim 78^\circ$. In
Fig.~\ref{fig:s}, we plot $S$ versus $\gamma$ (and $\alpha$), where
we parameterize $V_{ub}=0.00368\, e^{-i\gamma}$. The best fitted
value is $\gamma=(87_{-7}^{+33})^\circ$, corresponding to
$\alpha=(71_{-33}^{+7})^\circ$, for $\beta=22^\circ$.

\subsubsection{$\overline B\to a_1(1260) \overline K$ decays}

The decays $\overline B\to a_1 \overline K$ are penguin-dominated.
Because the dominant axial-vector and pseudoscalar penguin coefficients, $a_4^p(a_1 \overline K)$ and $a_6^p(a_1 \overline K)$, are constructive in the $a_1 \overline K$ modes,  $\overline B \to a_1 \overline K$ and the corresponding $\overline B\to \pi \overline K$ decays should have similar rates. It is instructive to consider the four ratios:
\begin{eqnarray}\label{eq:r-1}
R_1 &=& \frac{{\cal B}(\overline B^0 \to a_1^+ K^-)}
  {{\cal B}(\overline B^0 \to \pi^+ K^-)} \nonumber\\
    &=&
 \Bigg(
 \frac{V_0^{Ba_1}(m_{K}^2)}{F_0^{B\pi}(m_{K}^2)}
 \Bigg)^2
 \Bigg(\frac{\alpha_4^c (a_1 \overline K)}{\alpha_4^c(\pi \overline K)}\Bigg)^2
 \nonumber\\
 & & \times
 \Bigg[1+ 2 \, {\rm Re}
 \Bigg(\frac{\beta_3^c (a_1 \overline K)-\frac{1}{2}\beta_{3,{\rm EW}}^c (a_1 \overline K)}
  {\alpha_4^c (a_1 \overline K)}
 -
  \frac{\beta_3^c (\pi \overline K)-\frac{1}{2}\beta_{3,{\rm EW}}^c (\pi \overline K)}{{\alpha_4^c (\pi \overline K)}}
  \Bigg)
  +\cdots  \Bigg]
 \,,\nonumber\\
R_2 &=& \frac{{\cal B}(B^- \to a_1^- \overline K^0)}
  {{\cal B}(B^- \to \pi^- \overline K^0)} \nonumber\\
    &=&
 \Bigg(
 \frac{V_0^{Ba_1}(m_{K}^2)}{F_0^{B\pi}(m_{K}^2)}
 \Bigg)^2
 \Bigg(\frac{\alpha_4^c (a_1 \overline K)}{\alpha_4^c(\pi \overline K)}\Bigg)^2
 \nonumber\\
 & & \times
 \Bigg[1+ 2 \, {\rm Re}
 \Bigg(\frac{\beta_3^c (a_1 \overline K)
 + \beta_{3,{\rm EW}}^c (a_1 \overline K)} {\alpha_4^c (a_1 \overline K)}
 -
  \frac{\beta_3^c (\pi \overline K)
  + \beta_{3,{\rm EW}}^c (\pi \overline K)}{{\alpha_4^c (\pi \overline K)}}
  \Bigg)
  +\cdots  \Bigg]
 \,,\nonumber\\
R_3 &=& \frac{{\cal B}(\overline B^0 \to a_1^0 \overline K^0)}
  {{\cal B}(\overline B^0 \to \pi^0 \overline K^0)}\nonumber \\
    &=&
 \Bigg(
 \frac{V_0^{Ba_1}(m_{K}^2)}{F_0^{B\pi}(m_{K}^2)}
 \Bigg)^2
 \Bigg(\frac{\alpha_4^c (a_1 \overline K)}
 {\alpha_4^c(\pi \overline K)}\Bigg)^2
 \Bigg[ 1 - 3  \, {\rm Re}\Bigg[
 \frac{\alpha_{3,{\rm EW}}^c (\overline K a_1)}{\alpha_4^c(a_1 \overline K)} r_1
 -
 \frac{\alpha_{3,{\rm EW}}^c (\overline K \pi)}
 {\alpha_4^c(\pi \overline K)} r_2 \Bigg]
 \nonumber\\
 &&  + 2 \, {\rm Re}
 \Bigg(\frac{\beta_3^c (a_1 \overline K)
 -\frac{1}{2}\beta_{3,{\rm EW}}^c (a_1 \overline K)}
  {\alpha_4^c (a_1 \overline K)}
 - \frac{\beta_3^c (\pi \overline K)
  -\frac{1}{2}\beta_{3,{\rm EW}}^c (\pi \overline K)}
  {{\alpha_4^c (\pi \overline K)}}
  \Bigg)
  +\cdots  \Bigg]
 \,,\nonumber\\
R_4 &=& \frac{{\cal B}(B^- \to a_1^0 K^-)}
  {{\cal B}(B^- \to \pi^0 K^-)} \nonumber\\
    &=&
 \Bigg(
 \frac{V_0^{Ba_1}(m_{K}^2)}{F_0^{B\pi}(m_{K}^2)}
 \Bigg)^2
 \Bigg(\frac{\alpha_4^c (a_1 \overline K)}
 {\alpha_4^c(\pi \overline K)}\Bigg)^2
 \Bigg[ 1 + 3  \, {\rm Re}\Bigg[
 \frac{\alpha_{3,{\rm EW}}^c (\overline K a_1)}
 {\alpha_4^c(a_1 \overline K)} r_1
 -
 \frac{\alpha_{3,{\rm EW}}^c (\overline K \pi)}
 {\alpha_4^c(\pi \overline K)} r_2 \Bigg]
 \nonumber\\
 &&  + 2 \, {\rm Re}
 \Bigg(\frac{\beta_3^c (a_1 \overline K)
 + \beta_{3,{\rm EW}}^c (a_1 \overline K)}
  {\alpha_4^c (a_1 \overline K)}
 -
  \frac{\beta_3^c (\pi \overline K)
  + \beta_{3,{\rm EW}}^c (\pi \overline K)}{{\alpha_4^c (\pi \overline K)}}
  \Bigg)
  +\cdots  \Bigg] \,,
 \end{eqnarray}
 where
 \begin{eqnarray}
 r_1 &=&\frac{F_0^{BK}(m_{a_1}^2) \, f_{a_1}}
           {V_0^{Ba_1}(m_{K}^2)\, f_K} \approx 1.9 \,, \\
 r_2 &=&\frac{\, F_0^{BK}(m_{\pi}^2) \, f_{\pi}}
           {F_0^{B\pi}(m_{K}^2 )\, f_K} \approx 1.1 \,,
 \end{eqnarray}
and the dots stand for the neglected terms which are numerically estimated to be
less than 1\% in magnitude. The ratios $R_{1,2,3,4}$, which are very
insensitive to $\gamma$, are approximately proportional to
$[V_0^{Ba_1}(m_{K}^2)/(F_0^{B\pi}(m_{K}^2)]^2$ and receive
corrections mainly from the electroweak penguin and annihilation
topologies. The value of the annihilation $\beta_3$ is sensitive to
$a_1^{\perp,a_1}$. The contributions originating from electroweak
penguin and annihilation amplitudes can be further explored by
taking into account the following measurements for ratios,
 \begin{eqnarray}
 \frac{R_1}{R_2} &\cong& 1 -3  \, {\rm Re}
 \Bigg(\frac{\beta_{3,{\rm EW}}^c (a_1 \overline K)}
  {\alpha_4^c (a_1 \overline K)}
 -
  \frac{\beta_{3,{\rm EW}}^c (\pi \overline K)}{{\alpha_4^c (\pi \overline K)}} \Bigg)\,,\label{eq:r-2}\\
  \frac{R_1}{R_2}-\frac{R_3}{R_4}  &\cong&  6  \, {\rm Re}\Bigg[
 \frac{\alpha_{3,{\rm EW}}^c (\overline K a_1)}{\alpha_4^c(a_1 \overline K)} r_1
 -
 \frac{\alpha_{3,{\rm EW}}^c (\overline K \pi)}{\alpha_4^c(\pi \overline K)} r_2 \Bigg]\,,\label{eq:r-3}\\
 \frac{R_1}{R_3} \cong \frac{R_4}{R_2}  &\cong& 1+ 3  \, {\rm Re}\Bigg[
 \frac{\alpha_{3,{\rm EW}}^c (\overline K a_1)}{\alpha_4^c(a_1 \overline K)} r_1
 -
 \frac{\alpha_{3,{\rm EW}}^c (\overline K \pi)}{\alpha_4^c(\pi \overline K)} r_2 \Bigg] \label{eq:sr}
 \nonumber\\
&\cong& 1 +\frac{1}{2}\Bigg(\frac{R_1}{R_2}-\frac{R_3}{R_4}\Bigg)
\,.
 \end{eqnarray}
Although the above ratios are parameterized according to the QCD
factorization, they can be treated in a model-independent way. It is
worth stressing that because $\Phi_\perp^{a_1}(u)$ is antisymmetric
under interchange of the quark and antiquark momentum fractions in
the SU(2) limit, the weak annihilations (and hard spectator
interactions), which could contribute sizable corrections to the
decay amplitudes, enter the $\overline B\to a_1 \overline K$
amplitude in a very different pattern compared with $\overline B\to
\pi \overline K$ decays. More relevant information about $X_A$ and
$a_1^{\perp, a_1}$ can thus be provided by the measurement of
$R_1/R_2$.

With default parameters, the direct CP asymmetries are analogous to the
corresponding $\overline B\to \pi \overline K$ modes; because
$A_{CP}$s are dominated by ${\rm Re}(V_{td}^* V_{tb}) \, {\rm
Im}(\alpha_4^c + \beta_3^c) \, {\rm Im}(V_{ud}^*(V_{ub})$ times
${\rm Re}[\alpha_1 +\alpha_2 F_1^{BK}f_{a_1}/(V_0^{Ba_1} f_K)]$ and
$-{\rm Re}[\alpha_2 F_1^{BK}f_{a_1}/(V_0^{Ba_1} f_K)]$ terms for
$a_1^0 K^-$ and $a_1^0 \overline K^0$ modes, respectively, their
direct CP asymmetries are thus a little larger than the
corresponding $\pi \overline K$ modes in magnitude due to the decay constant
enhancement. Note that the value of $\beta_3$ is sensitive to the
first Gegenbauer moment of $\Phi_\perp^{a_1}(u)$ and the annihilation parameters
$\rho_A$ and $\phi_A$.
On the other hand, an outstanding problem is the determination of
the signs for direct CP observations in the $\pi \overline K$ modes. The
experimental results are $A_{CP}(\overline B^0 \to \pi^+ K^-)=
-0.095\pm 0.013$ and $A_{CP}(B^- \to \pi^0 K^-)= 0.046\pm 0.026$
\cite{hfag}. Some proposals, for instance the contribution due to
new-physics in the SM electroweak penguin
sector~\cite{Buras:2003dj,Buras:2004ub,Baek:2007yy} or due to
FSIs~\cite{Chua:2002wk,Cheng:2004ru}, were advocated for the
resolution. The ratio measurements for $R_1/R_2 -R_3/R_4, R_1/R_3$,
and $R_4/R_2$ directly probe the electroweak penguins. Moreover, the
approximate relation given in Eq.~(\ref{eq:sr}) will be violated if
the FSI patterns are different between $a_1 \overline K$ and $\pi \overline K$ modes.

\section{Conclusions}

We have studied $\overline B\to a_1(1260)\, \pi, a_1(1260) \overline
K$ decays. This paper is the first one in the literature using the QCD factorization approach
to study $B\to AP$ decays. Interestingly, due
to the G-parity, the leading-twist LCDA $\Phi_\perp^{a_1}$ of the
$a_1(1260)$ defined by the nonlocal tensor current is antisymmetric
under the exchange of $quark$ and $anti$-$quark$ momentum fractions
in the SU(2) limit, whereas the $\Phi_\parallel^{a_1}$ defined by
the nonlocal axial-vector current is symmetric. The large magnitude
of the first Gegenbauer moment ($a_1^{\perp,a_1}$) of $\Phi_\perp^{a_1}$ could have a
sizable impact on the annihilation amplitudes. If one ignores $\Phi_\perp^{a_1}$, i.e.,
letting $a_1^{\perp,a_1}=0$, with default parameters (where $\rho_A=0$), the branching ratio
for $a_1^0 \overline K^0$ mode becomes 1.8 times smaller,
while the changes of branching ratios for $a_1 \pi$ and the remaining $a_1 \overline K$ modes
are at the level of 5\% and 10\%, respectively.

Our main results are
summarized as follows.
\begin{itemize}
\item
Our results for ${\cal B}(\overline B^0 \to a_1^+ \pi^-,
a_1^- \pi^+)$  are in good agreement
with the data. Theoretically, the rates for $\overline B\to a_1(1260)\, \pi$
are close to the corresponding ones for $\overline B\to \rho \,
\pi$. The differences between the above two modes are mainly caused
by different magnitudes of form factors ($V_0^{Ba_1}$ and
$A_0^{B\rho}$) and decay constants ($f_{a_1}$ and $f_\rho$), and by
different patterns of tree--penguin interference.
For $\sin\gamma>0$, the T-P
interference is destructive in $\overline B^0\to a_1^\mp\pi^\pm, B^-
\to a_1^0\pi^-$, but constructive in $B^- \to a_1^-\pi^0$.
Because the amplitudes of $a_1 \pi$ and $\rho \pi$ modes are dominated by terms
with $\alpha_{1}$ and $\alpha^p_{4}$, and ${\rm Re}[\alpha_4^p(\pi
a_1)] \approx {\rm Re}[\alpha_4^p(a_1 \pi)/3] \approx {\rm
Re}[\alpha_4^p(\pi \rho)] \approx - {\rm Re}[\alpha_4^p(\rho \pi)]
\approx -0.034$, we obtain the relations as given in Eqs.~(\ref{eq:ratio-1})
and (\ref{eq:ratio-2}). Thus estimates for form factors and decay
constants as well as the weak phase $\gamma$ can thus be made
from these ratio measurements.

\item For $CP$ asymmetries, the large experimental errors do not allow us
to draw any particular conclusion in comparison with theoretical predictions.
The time-dependent CP asymmetry measurement in $B^0\to a_1^\pm
\pi^\mp$ can lead to the accurate determination of the CKM angle
$\gamma$. Using the current fitted value $\gamma=59.0^\circ$, i.e.,
$\alpha=99.0^\circ$ corresponding to $\beta=22.0^\circ$ in the SM,
our results show that $S$ and $\alpha_{\rm eff}$ differ from the present
data at about the $3.7\sigma$ level. This puzzle may
be resolved by using a larger $\gamma \gtrsim 80^\circ$. Further
measurements can clarify this discrepancy.

\item The branching ratios for the decays $B\to a_1 \pi$ and $a_1 K$ are
highly sensitive to the magnitude of $V_0^{Ba_1}(0)$. Using the LC sum rule result, $V_0^{Ba_1}(0)=0.28\pm 0.03$ \cite{kcy:FF}, the resultant branching ratios for $a_1^\pm \pi^\mp$ modes consist with the data very well. Nevertheless, the value of $V_0^{Ba_1}(0)$ is about 0.13 and $1.02\sim 1.22$ in the quark model calculations in Ref.~\cite{Cheng:2003sm} and Refs.~\cite{Scora:1995ty,Deandrea:1998ww}, respectively. If the quark model result is used in the calculation, ${\cal B}(\overline B^0\to a_1^\pm \pi^\mp)$ will be too small or large as compared with the data.

\item The $\overline B\to a_1 \overline K$ amplitudes resemble the corresponding $\overline B\to \pi \overline K$ amplitudes very much. Taking the ratios of corresponding CP-averaged $a_1 \overline K$ and $\pi \overline K$ branching ratios, we can extract information about the transition form factors, decay constants, electroweak penguin ($\alpha_{3,{\rm EW}}^c(\overline Ka_1)$), and annihilation topology ($\beta_{3,{\rm EW}}^c(a_1 \overline K)$). See Eqs.~(\ref{eq:r-2})-(\ref{eq:sr}). Thus, the possibilities for existing new-physics in the electroweak penguin sector and for final state interactions during decays can be explored.
\end{itemize}

\vskip0.2cm
{\it Note added.} Recently Belle has updated the following measurement \cite{:2007jn}:
${\cal B}(\overline B^0 \to a_1^+\pi^- + a_1^- \pi^+)=(29.8\pm3.2\pm 4.6)\times 10^{-6}$ which is in good
agreement with our result. On the other hand, BaBar has reported new measurements on
$a_1^0 \pi^-$, $a_1^-\pi^0$ and $a_1^+K^-$, $a_1^- \overline K^0$ modes \cite{:2007kp,Brown},
where ${\cal B}(\overline B^0\to a_1^+K^-)=(16.3\pm2.9\pm2.3)\times 10^{-6}$ is also in good agreement
with our prediction,  whereas the central values of branching ratios for the remaining modes are about
$2\sim 3$ times larger than our predictions. The latter discrepancies should be clarified
by the improved measurements in the future.

\begin{acknowledgments}
 I am grateful to H.~Y.~Cheng for useful comments.  This work was
supported in part by the National Science Council of R.O.C. under
Grant No: NSC95-2112-M-033-001.
\end{acknowledgments}

\vskip1cm

\appendix

\section{The coefficients $a_i^p$}\label{app:ai}
In the below discussion, we set $\Phi_\parallel^P \equiv \Phi^P$. In
Eq.~(\ref{eq:ai}), the expressions for effective parameters
$a_i^{p}$ are
 \begin{eqnarray}\label{eq:app-ai}
 a_i^{p}(M_{1} M_{2})\! &=& \! \bigg(c_i+{c_{i\pm1}\over
 N_c}\bigg)N_i(M_2) \nonumber\\
 && +{c_{i\pm1}\over N_c}\,{C_F\alpha_s\over
 4\pi}\Big[V_i(M_{2})+{4\pi^2\over
 N_c}H_i(M_{1} M_{2})\Big]+P_i^{p}(M_{2}).{\hspace{0.5cm}}
 \end{eqnarray}
 $N_i$ is given in Eq.~(\ref{eq:loterms}). The vertex corrections have the same expressions as
those for $VP$ modes \cite{BN} with LCDAs of the vector meson being replaced by the
corresponding ones of the $a_1$ meson. For the penguin contractions $P_i^p(M_2)$, one can
perform the same replacements but needs to add an overall minus sign to $P_6^p(a_1)$ and $P_8^p(a_1)$.
$H_i(M_1 M_2)$ have the expressions:
\begin{eqnarray}\label{eq:sepc01}
  H_i(M_1 M_2) &=& {-if_B f_{M_1} f_{M_2} \over X^{(\overline{B} M_1, M_2)}}
  \int^1_0 d\rho {\Phi^B_1(\rho)\over 1-\rho}\nonumber\\
   & & \times \int^1_0 d v \int^1_0 d u \,
 \Bigg( \frac{\Phi^{M_1}_\parallel(v) \Phi^{M_2}_\parallel(u)}{\bar u \bar v}
 \pm r_\chi^{M_1}
  \frac{\Phi_{m_{1}} (v) \Phi^{M_2}_\parallel(u)}{u \bar v}\Bigg),
 \hspace{0.5cm}
 \end{eqnarray}
for $i=1-4,9,10$,
\begin{eqnarray}\label{eq:spec02}
  H_i(M_1 M_2) &=& {if_B f_{M_1} f_{M_2}  \over X^{(\overline{B} M_1,
  M_2)}}
 \int^1_0 d\rho {\Phi^B_1(\rho)\over 1-\rho} \nonumber\\
 & &\times \int^1_0 d v \int^1_0 d u \,
 \Bigg( \frac{\Phi^{M_1}_\parallel(v) \Phi^{M_2}_\parallel(u)}{u \bar v}
 \pm  r_\chi^{M_1}
  \frac{\Phi_{m_1} (v) \Phi^{M_2}_\parallel (u)}{\bar u \bar v}\Bigg),
 \hspace{0.5cm}
 \end{eqnarray}
for $i=5,7$, and $H_i(M_1 M_2)=0$ for $i=6,8$, where the upper
(lower) signs apply when $M_1=P$ ($M_1=A$). Here $\Phi^B_1(\rho)$ is
one of the two LCDAs of the $\overline
B$ meson~\cite{BBNS}.

 \section{The annihilation amplitudes $A_n^{i,f}$}\label{app:A}

For $A_n^{i,f}$ (see Eq.~(\ref{eq:bi})), some signs change in comparison with the results of $B\to PP$ and $PV$.
We obtain
\begin{eqnarray}\label{blocks}
   A_1^i &=& \pi\alpha_s \int_0^1\! dx dy\,
    \left\{ \Phi_\parallel^{M_2}(x)\,\Phi_\parallel^{M_1}(y)
    \left[ \frac{1}{y(1-x\bar y)} + \frac{1}{\bar x^2 y} \right]
    - r_\chi^{M_1} r_\chi^{M_2}\,\Phi_{m_2}(x)\,\Phi_{m_1}(y)\,
     \frac{2}{\bar x y} \right\} ,
    \nonumber\\
   A_1^f &=& A_2^f = 0 \,, \nonumber\\
   A_2^i &=& \pi\alpha_s \int_0^1\! dx dy\,
    \left\{ \Phi_\parallel^{M_2}(x)\,\Phi_\parallel^{M_1}(y)
    \left[ \frac{1}{\bar x(1-x\bar y)} + \frac{1}{\bar x y^2} \right]
    - r_\chi^{M_1} r_\chi^{M_2}\,\Phi_{m_2}(x)\,\Phi_{m_1}(y)\,
     \frac{2}{\bar x y} \right\} ,
    \nonumber\\
   A_3^i &=& \pm \pi\alpha_s \int_0^1\! dx dy\,
    \left\{r_\chi^{M_1}\,\Phi_\parallel^{M_2}(x)\,\Phi_{m_1}(y)\,
    \frac{2\bar y}{\bar x y(1-x\bar y)}
    + r_\chi^{M_2}\,\Phi_\parallel^{M_1}(y)\,\Phi_{m_2}(x)\,
    \frac{2x}{\bar x y(1-x\bar y)} \right\} , \nonumber\\
   A_3^f &=& \pm \pi\alpha_s \int_0^1\! dx dy\,
    \left\{r_\chi^{M_1}\,\Phi_\parallel^{M_2}(x)\,\Phi_{m_1}(y)\,
    \frac{2(1+\bar x)}{\bar x^2 y}
    -  r_\chi^{M_2}\,\Phi_\parallel^{M_1}(y)\,\Phi_{m_2}(x)\,
    \frac{2(1+y)}{\bar x y^2} \right\} ,
\end{eqnarray}
where the upper (lower) signs apply when $M_1=P$ ($M_1=A$) and the
detailed definitions of the distribution amplitudes of the
axial-mesons have been collected in Sec.~\ref{sec:DA}. Again, here
we have set $\Phi_\parallel^P \equiv \Phi^P$.

Using the asymptotic distribution amplitudes of
$\Phi_\parallel^{a_1}(u)$ and $\Phi_P(u)$, and the approximation
$\Phi_\perp^{a_1}(u)=18 u\bar u  (2u-1) a_1^{\perp,a_1}$, we obtain
the annihilation amplitudes
\begin{eqnarray}
 A_1^i &\approx &  6\pi\alpha_s \left[\,
    3\,\bigg( X_A - 4 + \frac{\pi^2}{3} \bigg)
    - a_1^{\perp,a_1} r_\chi^{a_1} r_\chi^{P} X_A (X_A -3) \right] , \\
 A_2^i &\approx &  6\pi\alpha_s \left[\,
    3\,\bigg( X_A - 4 + \frac{\pi^2}{3} \bigg)
    - a_1^{\perp,a_1} r_\chi^{a_1} r_\chi^{P} X_A (X_A -3) \right] , \\
 A_3^i &\approx & \pm 6\pi\alpha_s\, \Bigg[ r_\chi^{P}
    \bigg( X_A^2 - 2 X_A + \frac{\pi^2}{3} \bigg)
    + 3 a_1^{\perp,a_1} r_\chi^{a_1} \Bigg(X_A^2 - 2X_A -6 +\frac{\pi^2}{3}\Bigg)
    \Bigg] \,, \\
 A_3^{f} & \approx& 6 \pi\alpha_s
(2 X_A -1) \bigg[ r_\chi^P X_A -
 3 a_1^{\perp, a_1} r_\chi^{a_1} (X_A-3) \bigg]\,.\label{eq:ann-A}
 \end{eqnarray}

\end{document}